\documentclass[aps,prb,onecolumn,superscriptaddress,floatfix]{revtex4-2}
\usepackage[utf8]{inputenc}
\usepackage{amsmath,amssymb,amsthm}
\usepackage{graphicx}
\usepackage{xcolor}
\usepackage{hyperref}
\usepackage{physics}
\usepackage{booktabs}
\usepackage{siunitx}
\usepackage{accents}

\begin{document}

\title{Stochastic Thermodynamics of Quantum-Induced Stochastic Dynamics}

\author{Pedro Ventura Paraguassú}
\affiliation{Departamento de F\'isica, PUC-Rio, 22452-970, Rio de Janeiro RJ, Brazil}

\date{\today}

\begin{abstract}
Quantum-Induced Stochastic Dynamics arises from the coupling between a classical system and a quantum environment. Unlike standard thermal reservoirs, this environment acts as a dynamic bath, capable of simultaneously exchanging heat and performing work. We formulate a thermodynamic framework for this semi-classical regime, defining heat, work, and entropy production. We derive a modified Second Law that accounts for non-equilibrium quantum features, such as squeezing. The framework is exemplified by an optomechanical setup, where we characterize the thermodynamics of the non-stationary noise induced by the cavity field.
\end{abstract}

\maketitle

\section{Introduction}

To understand behavior at microscopic scales, quantum dynamics was developed. While this framework successfully predicts the behavior of atoms, electrons, and other fundamental constituents, the theory itself does not explicitly prescribe the limits of its validity. Current experiments aim to probe the boundaries where quantum dynamics can be observed, including on levitated nanoparticles \cite{kremer2024all,tebbenjohanns2021quantum, delic2020cooling, piotrowski2023simultaneous} and clamped optomechanical systems \cite{o2010quantum, chan2011laser, bild2023schrodinger, ockeloen2018stabilized}. Although these systems are typically described by classical mechanics, they exhibit quantum behaviors due to their high degree of isolation or their interaction with inherently quantum systems.

In such scenarios, a classical stochastic dynamics can emerge from the interaction with a quantum system. Recently, this regime has been investigated in optomechanics \cite{paraguassu2024quantum, paraguassu2025apparent}, the Jaynes-Cummings interaction \cite{sobrero2025response}, and even in gravitational wave detection \cite{parikh2021signatures, parikh2025quantum}. We refer to this as Quantum-Induced Stochastic Dynamics (QISD), where stochasticity arises solely from the interaction with a quantum degree of freedom \cite{caldeira1983path, calzetta1994noise, weiss2012quantum, kamenev2023field}. Here, the effect of the quantum system manifests as fluctuation and dissipation, effectively acting as a reservoir that exchanges energy with the classical system.  By considering a strong decoherence limit on the system of interest, the dynamics becomes a generalized Langevin equation characterized by a Wiener measure \cite{camurati2026}. It is different from quantum Brownian motion, since the Brownian particle is classical \cite{artini2025non, caldeira1983path}.

Given the stochastic nature of the dynamics, the energy exchange is inherently stochastic, making the framework of stochastic thermodynamics well-suited to investigate this exchange from the perspective of the system's dynamics. Stochastic thermodynamics treats thermodynamic functionals—such as heat, work, and entropy—as random variables. The success of the theory relies on extending the second law of thermodynamics through fluctuation theorems \cite{seifert2012stochastic, jarzynski2011equalities, sekimoto2010stochastic, peliti2021stochastic, seifert2025stochastic}, which find applications ranging from RNA measurements \cite{collin2005verification, alemany2012experimental, manosas2007force} to Brownian machines \cite{martinez2016brownian, holubec2021fluctuations, li2024realization, forao2025statistics}. Originally established by the works of Sekimoto \cite{sekimoto1998langevin}, Seifert \cite{seifert2005entropy}, and Jarzynski \cite{jarzynski1997nonequilibrium} almost three decades ago, the framework has since seen various generalizations, some of them includes special relativistic \cite{paraguassu2021heat, pal2020stochastic, koide2011thermodynamic} and general relativistic extensions \cite{tao2024general, cai2025fluctuation, cai2023relativistic, cai2023relativistic2, pei2025promoting}, active matter \cite{caprini2019entropy, bebon2025thermodynamics, dabelow2019irreversibility, paraguassu2025effects}, viscoelastic bath \cite{darabi2023stochastic, guevara2023brownian}, computing \cite{wolpert2024stochastic_perspective, wolpert2019stochastic, helms2025stochastic}, opinion dynamics \cite{tome2023stochastic, hawthorne2023nonequilibrium}, quantum systems \cite{funo2018path, binder2015quantum, strasberg2019operational, elouard2017role}, and quantum fields \cite{koide2025unification}. In this work, we propose to understand the exchange of energy of the QISD dynamics, 
thereby incorporating these dynamics into the broader context of stochastic thermodynamics.

Our objective is to employ the framework of stochastic thermodynamics to characterize the energy exchanges within Quantum-Induced Stochastic Dynamics. We seek to elucidate, from a semiclassical perspective, the fundamental mechanisms of energy transfer between a probe and its quantum environment. Comprehending these exchanges at the classical-quantum interface is essential for the development of protocols designed to harvest or utilize quantum-induced energy. We find that the quantum reservoir acts as a dynamic bath, capable of simultaneously exchanging heat and performing work, thereby challenging the standard tripartite separation in stochastic thermodynamics. Consequently, the Second Law of Thermodynamics is generalized through path irreversibility, establishing a precise connection between heat dissipation, entropy production, and the non-equilibrium resources provided by squeezing.

The paper is organized as follows: In Section 2, we define QISD, demonstrating through the Feynman-Vernon formalism how stochastic dynamics naturally emerges. In Section 3, we discuss the definitions of work and heat, addressing the distinction between them given the ultimately quantum nature of the forces involved. In Section 4, we propose a formulation for entropy production based on path entropy to quantify the system's irreversibility; we show that if the fluctuation-dissipation theorem (FDT) holds, irreversibility can be directly connected to heat exchange. Finally, in Section 5, we apply this framework to the example of a trapped nanoparticle exhibiting stochastic behavior due to its interaction with squeezed light modes in a cavity. We conclude with a discussion of the results and future perspectives.

\section{Quantum-Induced Stochastic Dynamics}

We start by considering a bipartite system composed of subsystems $A$ and $B$. In our framework, we treat $A$ as the classical system of interest, while $B$ represents the quantum degrees of freedom. The time evolution of the joint density matrix is given by $\hat{\rho}_{\text{tot}}(t) = \hat{U}(t) \hat{\rho}_0 \hat{U}^\dagger(t)$. We obtain the reduced density matrix of system $A$ by taking the partial trace over the degrees of freedom of $B$, assuming it is prepared in a separable quantum state, and representing the result in the position basis $\{\ket{q}\}$ of system $A$. By expressing the time-evolution operator in terms of path integrals, we arrive at the Feynman-Vernon functional formalism~\cite{feynman2000theory}
\begin{equation}
    \rho(q, q', t) = \int dq_0 dq'_0 \, \mathcal{J}[q, q'; q_0, q'_0] \, \rho_0(q_0, q'_0).
\end{equation}
The central object of our analysis is the propagator of the reduced density matrix, $\mathcal{J}$, which accounts for the free evolution of the system modified by the coupling to the environment via the influence functional
\begin{equation}
    \mathcal{J}[q, q'; q_0, q'_0] = \int \mathcal{D}q \int \mathcal{D}q' \, e^{\frac{i}{\hbar}(S_A[q] - S_A[q'])}\mathcal{F}[q,q'],
\end{equation}
where $S_A$ is the classical action of the system $A$, defined by $S_A[q]=\int_0^t d\tau (\frac{1}{2}m \dot{q}^2 - V(q))$, and $\mathcal{F}$ is the influence functional encapsulating the effects of the bath.

Assuming an interaction Hamiltonian  $H_{int} = -f(q) \hat{B}$ (a generalization is provided in Ref.~\cite{camurati2026}), where $\hat{B}$ is a bath operator depending on the quadrature $Q$ and $f(q)$ an arbitrary function. If we keep only quadratic terms in the functional, we will have a Gaussian influence functional given by~\cite{calzetta1994noise}
\begin{align} \label{eq:influence_functional}
    \mathcal{F}[q,q'] = \exp\Bigg\{ &-\frac{1}{\hbar} \int_0^\tau dt \int_0^\tau dt' \, [q(t)-q'(t)] K(t,t') [q(t')-q'(t')] \nonumber \\
    &+ \frac{i}{\hbar} \int_0^\tau dt \int_0^t dt' \, [q(t)-q'(t)] D(t, t') [\dot q(t')+\dot q'(t')] \Bigg\}.
\end{align}
Here, the real part of the exponent is associated with decoherence (noise), determined by the noise kernel $K$, while the imaginary part drives the dissipation via the dissipation kernel $D$. The effective dynamics are best analyzed by introducing the center-of-mass coordinate $x(t) = (q+q')/2$ and the difference (coherence) coordinate $y(t) = q-q'$~\cite{weiss2012quantum,funo2018path}. In these variables, Eq.~(\ref{eq:influence_functional}) takes the form
\begin{equation}
    \mathcal{F}[x,y] = \exp\left\{ -\frac{1}{\hbar} \int_0^\tau dt \int_0^\tau dt' \, y(t) K(t,t') y(t') + \frac{i}{\hbar} \int_0^\tau dt \int_0^t dt' \, y(t) D(t, t') \dot x(t') \right\}.
\end{equation}
Depending on the initial state of the bath, additional deterministic terms may arise, acting as driving forces on the classical system - for instance, when the bath is prepared in a coherent state~\cite{paraguassu2024quantum, parikh2021signatures}. In such cases, a deterministic force $F_{\rm det}(t)$ acts on the system $A$, manifesting as an additional linear term in the effective action: $\frac{i}{\hbar}\int_0^\tau dt \, F_{\rm det}(t) y(t)$.

The classical limit emerges when we assume that the coherence length $y$ is negligible compared to the characteristic scales of the problem. This assumption collapses the quantum Feynman path integral into a stochastic Wiener path integral~\cite{camurati2026}. By employing the Wigner representation of the density matrix, we arrive at the phase-space probability distribution evolution:
\begin{equation}
    W(x,p, t) = \int dx_0 dp_0 \, W(x_0,p_0, 0) \, P[x,p; x_0,p_0],
\end{equation}
where the conditional probability distribution is governed by the Wiener measure:
\begin{equation}
    P[x,p; x_0,p_0] = \int \mathcal{D}x \, \exp\left\{ -\frac{1}{2} \int_0^\tau dt \int_0^\tau dt' \, \mathcal{E}[x(t)] K^{-1}(t,t') \mathcal{E}[x(t')] \right\},
\end{equation}
with the dynamical equation functional defined as $\mathcal{E}[x] = m \ddot{x} + V'(x) - F_{\text{diss}}-F_{\rm det}(t)$. This formulation reveals that the system evolves according to a Generalized Langevin Equation (GLE) \cite{zwanzig2001nonequilibrium}
\begin{equation}
    m \ddot{x}(t) + \partial_x V(x) = F_{\text{diss}}(t) + \eta(t)+F_{\rm det}(t),\label{eq:Langevin}
\end{equation}
where $\eta(t)$ is a Gaussian colored noise with zero mean and correlation function given by the noise kernel
\begin{equation}
    \langle \eta(t) \eta(t') \rangle = K(t,t').
\end{equation}
and the dissipation force, that appears from influence functional, is
\begin{equation}
    F_{\rm diss}(t) = - \int_0^\tau D(t,t') \dot x dt'.
\end{equation}

While the generalized Langevin equation appears in various classical contexts, its origin here is explicitly traced back to the influence functional. The noise term arises from the decoherence mechanism, while the dissipation (which removes energy from the system) represents the backreaction of the quantum environment \cite{calzetta1994noise}. Understanding these dynamical effects is essential for analyzing the system's energetics, as we discuss next.

\section{Energetics of Heat and Work}

Based on the effective Langevin dynamics derived in Eq.~(\ref{eq:Langevin}), we now address the system's energetics, starting with the definition of stochastic work. In scenarios where the reservoir acts as a source of a deterministic driving force $F_{\text{det}}(t)$---for instance, when prepared in a coherent state---it induces controlled motion in the system. Treating this force as an external drive, we define the work as
\begin{equation}\label{eq:work def}
   W[x] = - \int_0^\tau dt \, {F}_{\text{det}}(t) \dot x(t).
\end{equation}
The negative sign adopted here follows the convention where positive work corresponds to energy transferred from the system to the environment (work done by the system). This definition aligns with the interpretation of a driving force discussed in Ref.~\cite{peliti2021stochastic}. An alternative definition considers the time derivative of the external force, aligning with the thermodynamic interpretation of work as the energy change resulting from the variation of an external control parameter. In such a framework, the force contributes to the free energy, which is particularly suitable for analyzing apparent energetic anomalies (or ``free lunches''). Indeed, work arising from a coherent state in this context has been previously proposed and investigated in Ref.~\cite{paraguassu2025apparent}.

Defining heat, however, requires careful reinterpretation. Conventionally, heat is identified as the energy exchanged with the thermal bath \cite{sekimoto2010stochastic}. In our setup, the quantum environment plays a threefold role: it acts simultaneously as the source of noise, dissipation, and the deterministic drive. To resolve this ambiguity, we distinguish the forces based on their capacity to induce ordered versus disordered motion. The noise term ($\eta$), arising from decoherence, induces random fluctuations. The dissipation term ($F_{\text{diss}}$), while unitary in origin, acts as a friction force that removes energy from the system. In contrast, $F_{\text{det}}$ provides a controlled, ordered energy input. Therefore, we associate heat with the energy exchange mediated by the non-conservative and stochastic forces
\begin{equation} \label{eq:heat_def}
    Q[x] = \int_0^\tau d\tau \, \dot{x}(t) \left[ F_{\text{diss}}(t) + \eta(t) \right].
\end{equation}

A direct consequence of this definition is the recovery of the first law of stochastic thermodynamics. By substituting the equation of motion $m \ddot{x} + \partial_x V = F_{\text{det}} + F_{\text{diss}} + \eta$ into Eq.~(\ref{eq:heat_def}), we obtain
\begin{align}
    Q[x] &= \int_0^\tau dt \, \dot{x} \left( m \ddot{x} + \partial_x V - F_{\text{det}} \right) \nonumber \\
         &= \int_0^\tau dt \frac{d}{dt} \left( \frac{m \dot{x}^2}{2} + V(x) \right) - \int_0^\tau dt \, F_{\text{det}} \dot{x} \nonumber \\
         &= \Delta K + \Delta V+ W[x] ,
\end{align}
which leads to the energy balance $\Delta \mathcal{E} = Q[x] - W[x]$. Here, $\mathcal{E} = K + V $ represents the total internal energy of the system, including the potential energy contribution from the driving force. Note that the variables $\Delta K$ and $\Delta V$ are also random variables having a probability distribution \cite{paraguassu2024brownian, melo2024brownian}.

This formulation highlights a crucial physical insight: the quantum environment operates as a dynamic bath. Unlike a passive reservoir that merely absorbs energy, it actively exchanges work with the system. This behavior is not exclusive to the quantum nature of the bath but rather to its specific preparation. For instance, if the quantum system is in a coherent state, work will be made \cite{paraguassu2025apparent}. An analogy can be drawn to classical experiments involving active feedback traps or electrical noise circuits \cite{jun2014high, pires2023optimal,sagawa2012nonequilibrium, martinez2016brownian, kremer2024all}, where the same apparatus can provide both random fluctuations (heat exchange) and deterministic control forces.

To ensure the thermodynamic consistency of these definitions, we must next examine the entropy production and its relation to the influence functional.

\section{Entropy Production and Fluctuation Theorem}

To validate the thermodynamic quantities derived in the previous section, we establish the connection between energetics and irreversibility. A standard approach involves computing the stochastic entropy production, defined as the logarithmic ratio of the forward and backward path probability densities. For standard thermal baths, this ratio recovers the usual entropy production, which is directly proportional to the dissipated heat. However, in more complex scenarios---such as active baths or non-equilibrium environments---deviations from this simple heat-entropy relation are expected~\cite{dabelow2019irreversibility}, signaling a breakdown of the fluctuation-dissipation relation in its simplest form.

The measure of irreversibility, identified as the path entropy production \cite{evans2002fluctuation, searles1999fluctuation}, is quantified by the logarithmic ratio of the forward and backward path probabilities
\begin{equation}
    \Delta \mathcal{I}[x] = \ln \frac{\mathcal{P}[x]}{\tilde{\mathcal{P}}[\tilde{x}]}.
\end{equation}
We analyze this relationship for our generalized Langevin dynamics. We begin by writing the probability weight for a forward trajectory $\{x(\tau)\}$ over the interval $t \in [0, \tau]$, which is governed by the Onsager-Machlup functional\cite{wio2013path}
\begin{equation}
    \mathcal{P}[\{x(t)\}_0^\tau] \propto \exp\left\{ -\frac{1}{2} \int_0^\tau dt \int_0^\tau dt' \, \mathcal{E}[x(t)] K^{-1}(t,t') \mathcal{E}[x(t')] \right\},
\end{equation}
The probability of the reverse process is determined by applying the time-reversal operation $\Theta$ (details in Appendix A). We define the time-reversed trajectory as $\tilde{x}(t) = x(\tau-t)$ for $t \in [0, \tau]$. Assuming that the statistical properties of the reservoir (encoded in the noise and dissipation kernels) remain invariant under time reversal, the probability weight for the backward path is given by
\begin{equation}
    \tilde{\mathcal{P}}[\{\tilde{x}(t)\}_0^\tau] \propto \exp\left\{ -\frac{1}{2} \int_0^\tau dt \int_0^\tau dt' \, \tilde{\mathcal{E}}[\tilde{x}(t)] K^{-1}(t,t') \tilde{\mathcal{E}}[\tilde{x}(t')] \right\},
\end{equation}
where the dynamical functional for the reverse path, taking into account the parity of the forces under time reversal, reads:
\begin{equation}
    \tilde{\mathcal{E}}[\tilde{x}] = m \ddot{\tilde{x}} + V'(\tilde{x}) - \tilde{F}_{\text{diss}}[\tilde{x}] - \tilde{F}_{\text{det}}(t).
\end{equation}
Here, $\tilde{F}_{\text{det}}(\tau) = F_{\text{det}}(\tau-t)$ represents the time-reversed driving protocol. As detailed in Appendix A, the dissipative force transforms oddly under time reversal due to its dependence on velocity ($\dot{\tilde{x}} = -\dot{x}$), leading to the relation $\tilde{F}_{\text{diss}}[\tilde{x}] = - F_{\text{diss}}[x]$.

Substituting the explicit forms of the probability functionals derived from the generalized Langevin dynamics, we obtain the expression for the path entropy production
\begin{equation} \label{eq:irreversibility}
    \Delta \mathcal{I} = 2 \int_0^\tau dt \int_0^\tau dt' \, F_N(t) \mathcal{K}(t,t') \dot{x}(t'),
\end{equation}
where $F_N(t) \equiv m \ddot{x} + V'(x) - F_{\text{det}}(t)$ represents the net force excluding the deterministic drive. Here, we have introduced the effective non-local kernel $\mathcal{K}$, defined by the convolution of the inverse noise kernel with the dissipation kernel
\begin{equation} \label{eq:superkernel}
    \mathcal{K}(t,t') = \int_{t'}^\tau ds \, K^{-1}(t,s) D(s,t').
\end{equation}
From the Langevin equation, we identify $F_N(t) = F_{\text{diss}}(t) + \eta(t)$. Consequently, Eq.~(\ref{eq:irreversibility}) bears a striking resemblance to the definition of heat in Eq.~(\ref{eq:heat_def}). It can be rewritten as
\begin{equation}
    \Delta \mathcal{I} = 2 \int_0^\tau dt \int_0^\tau dt' \, \left[F_{\rm diss}(t) + \eta(t) \right] \mathcal{K}(t,t') \dot{x}(t').
\end{equation}
This expression suggests a generalized mechanism of energy exchange between the reduced system and the quantum environment. However, unlike the standard definition of heat, the forces here are coupled to the velocity via the non-local kernel $\mathcal{K}$. Simply defining this quantity as a `generalized heat' is insufficient, as it does not automatically guarantee validity of the second law of thermodynamics without further constraints on the kernel structure.

We can proceed by considering the regime where the environment satisfies the FDT. Mathematically, this implies that the inverse noise kernel acts as the inverse of the dissipation kernel (up to a constant factor), such that
\begin{equation}
    \int_0^\tau ds \, K^{-1}(t,s) D(s,t') \propto \delta(t - t').
\end{equation}
For Eq.~(\ref{eq:superkernel}) to reduce to a delta function strictly, the dissipation must be causal (requiring $s-t' \le 0$ within the integration limits). When this condition holds, the inverse noise kernel effectively `undoes' the memory effects of the dissipation. In this limit, $\mathcal{K}(t, t') \to \mathcal{C}\delta(t-t')$ (where $\mathcal{C}$ is a constant, that could be the inverse temperature, $\beta$, depending on the state), and the irreversibility becomes directly proportional to the heat exchanged, $\Delta \mathcal{I} \approx \mathcal{C} Q[x]$. This recovers the standard stochastic thermodynamics result, where irreversibility is identified with the entropy produced in the medium \cite{cohen2008properties, chernyak2006path}.

We now generalize this framework to scenarios involving an additional noise source that violates the standard FDT; for instance, when the quantum enviroment is prepared in a squeezed state, introducing non-stationary fluctuations~\cite{paraguassu2024quantum, parikh2021signatures}. We model the total noise kernel as
\begin{equation}
    K(t,s) = K_0(t,s) + \mathcal{V}(t,s).
\end{equation}
where $K_0$ satisfies the FDT and $\mathcal{V}$ could represents a non-stationary contribution or deviation. Assuming $\mathcal{V}$ is associated with a small parameter, we invert the kernel using the Neumann expansion \cite{itzykson2006quantum}, a procedure analogous to the Dyson series expansion used in quantum field theory to obtain the full propagator
\begin{equation}
    K^{-1}(t,s) \approx K_0^{-1}(t,s) - \int_0^\tau du \int_0^\tau dv \, K_0^{-1}(t,u) \mathcal{V}(u,v) K_0^{-1}(v,s).
\end{equation}
Substituting this expansion into Eq.~(\ref{eq:superkernel}), the effective kernel becomes:
\begin{equation}
    \mathcal{K}(t,t') \approx \mathcal{C} \delta(t-t') - \mathcal{C} \int_0^\tau du \, \mathcal{V}(u,t) K_0^{-1}(u,t').
\end{equation}
Finally, returning to the irreversibility expression, we find:
\begin{equation}
    \Delta \mathcal{I} \approx 2\mathcal{C}Q[x] - 2\mathcal{C} \int_0^\tau dt \int_0^\tau dt' \int_0^\tau du \, F_N(t) \left[ \mathcal{V}(u,t)K_0^{-1}(u,t') \right] \dot{x}(t').
\end{equation}
A deviation from the FDT gives rise to an additional term in the total entropy production. This correction accounts for the irreversibility introduced by the non-equilibrium preparation of the quantum system, which typically involves non-classical states \cite{sobrero2025response, paraguassu2024quantum}. While the physical interpretation of this term is system-dependent, we illustrate its significance in the next section with an example where it appears due to squeezing.

With the definition of irreversibility established, we now address the validity of the fluctuation theorem in this framework. The Integral Fluctuation Theorem (IFT) for the total entropy production is satisfied by construction, as the total entropy change along a trajectory is defined by the logarithmic ratio of path probabilities \cite{seifert2005entropy}
\begin{equation}
    \Delta s_{\text{tot}}[x] = \ln\frac{p_0(x_0)\mathcal{P}[x]}{\tilde{p}_0(\tilde{x}_0)\tilde{\mathcal{P}}[\tilde{x}]} = \ln\frac{p_0(x_0)}{\tilde{p}_0(\tilde{x}_0)} + \ln\frac{\mathcal{P}[x]}{\tilde{\mathcal{P}}[\tilde{x}]}.
\end{equation}
Here, $p_0(x_0)$ is the initial distribution of the forward process. For the standard IFT to hold, the initial distribution of the reverse process, $\tilde{p}_0$, is chosen to be the final distribution of the forward process, i.e., $\tilde{p}_0(\tilde{x}_0) = p_t(x_t)$. The first term on the right-hand side corresponds to the change in stochastic entropy of the system, $\Delta s_{\text{sys}} = -\ln p_t(x_t) + \ln p_0(x_0)$~\cite{seifert2005entropy}, while the second term is the path irreversibility $\Delta \mathcal{I}$ derived previously. By definition, the ensemble average of the exponential of the total entropy satisfies
\begin{equation}
    \langle e^{-\Delta s_{\text{tot}}} \rangle = 1.
\end{equation}
Applying Jensen's inequality, $\langle e^{-X} \rangle \ge e^{-\langle X \rangle}$, implies the non-negativity of the mean total entropy production, $\langle \Delta s_{\text{tot}} \rangle \geq 0$. Substituting the expression for $\Delta \mathcal{I}$ obtained in the previous section, we arrive at the modified Second Law for the Quantum-Induced Stochastic Dynamics
\begin{equation}
    \langle \Delta s_{\text{sys}} \rangle + 2\mathcal{C} \langle Q[x] \rangle \geq 2\mathcal{C}\int_0^\tau dt \int_0^\tau dt'  \, \langle F_N(t) \dot{x}(t')\rangle \int_0^\tau du \mathcal{V}(u,t)K_0^{-1}(u,t') .
\end{equation}
This result establishes a new thermodynamic bound due to deviations from the FDT. The term on the right-hand side combines the contribution of the additional non-stationary noise (associated, for instance, with squeezing) with the dissipation. Consequently, the standard entropy production (system plus heat) is no longer strictly positive but is bounded by the noise correlations injected by the quantum system.

By connecting the path entropy with the heat exchanged, we demonstrate the internal consistency of the thermodynamic interpretation. It is worth noting, however, that the explicit form of this bound assumes the validity of the FDT and causality for the unperturbed part of the kernel ($K_0$). We proceed to illustrate these concepts with specific a example.

\section{Stochastic Thermodynamics of a nanoparticle interacting with quantum light}

To illustrate the developed framework, we consider a concrete optomechanical setup consisting of a levitated nanoparticle interacting with a cavity field, as described in Ref.~\cite{delic2019cavity}. The Hamiltonian, assuming a single optical mode and the coherent scattering approximation, is given by
\begin{equation}
    H/\hbar = \omega_c a^\dagger a + \omega_m b^\dagger b + g (a^\dagger+a) (b^\dagger + b).
\end{equation}
Here, the mechanical motion of the nanoparticle constitutes the system of interest (represented by the operators $b, b^\dagger$, whose classical limit corresponds to the position $x$ and momentum $p$), while the cavity field acts as the quantum environment (represented by $a, a^\dagger$). Assuming a Lorentzian spectral density for the cavity field, $J(\omega) = \frac{1}{\pi} \frac{\gamma}{(\omega - \omega_c)^2 + \gamma^2}$~\cite{fox2006quantum}, and applying the Feynman-Vernon formalism, we derive the effective QISD equation for the particle
\begin{equation}
    m \ddot{x}(t) + kx = F_{\text{diss}}(t) + \eta(t) + F_{\text{det}}(t),
\end{equation}
where we neglect the contribution from the classical environment that can be ignored in a time interval of $\sim 1ms$ and the recoil heating, that could be suppressed by changing the trap \cite{gonzalez2023suppressing}. This system serves as an ideal testbed, as its dynamics have been extensively analyzed in Ref.~\cite{paraguassu2024quantum}. It is well-established that for a thermal state of the cavity, the FDT holds. However, for a displaced squeezed thermal state, 
\begin{equation}
    \hat\rho_{0} = \hat D(\alpha) \hat S(r) \hat\rho_{\text{th}} \hat{S}^\dagger(r) \hat{D}^\dagger(\alpha),
\end{equation} 
where $D(\alpha)$ is the displacement operator (generating the coherent component), $S(r)$ is the squeezing operator, with $r$ being the squeezing radius, and $\rho_{\text{th}}$ is the thermal density matrix, the FDT is violated due to the emergence of non-stationary components. Specifically, the noise can be decomposed into stationary and non-stationary parts, $\eta(t) = \eta_{\text{st}}(t) + \eta_{\text{n-st}}(t)$, with correlation functions given by
\begin{align}
    \langle \eta_{\text{st}}(t) \eta_{\text{st}}(t') \rangle &= 2 \cosh(2r) \left( \frac{k_B T}{\hbar \omega_c} \right) F_0^2 \, e^{-\gamma|t-t'|} \cos[\omega_c (t-t')], \\
    \langle \eta_{\text{n-st}}(t) \eta_{\text{n-st}}(t') \rangle &= 2 \sinh(2r) \left( \frac{k_B T}{\hbar \omega_c} \right) F_0^2 \, e^{-\gamma(t+t')} \cos[\omega_c(t+t')],
\end{align}
where $F_0 = g \sqrt{2 \hbar m \omega_m}$ represents the effective force intensity associated with the coupling $g$, and $r$ is the squeezing parameter, typically is a small force about $10^{-18}N$ \cite{sobrero2025response, paraguassu2024quantum}. The deterministic driving force $F_{\rm det}(t)$, arising from the coherent displacement $\alpha$, is given by
\begin{equation}
    F_{\rm det}(t) = 2 F_0 |\alpha| e^{-(\gamma t+r)} \cos(\omega_c t),
\end{equation}
while the dissipative force, governed by the time-delayed response of the cavity, reads
\begin{align}
    F_{\text{diss}}(t) &= - \frac{F_0^2}{\hbar \omega_c} \int_0^t dt' \, \dot{x}(t) e^{-\gamma |t-t'|} \cos[\omega_c (t-t')] \equiv - \int_0^t dt' \, \dot{x}(t') D(t,t'), \\
    D(t,t') &= \frac{F_0^2}{\hbar \omega_c} e^{-\gamma |t-t'|} \cos[\omega_c (t-t')].
\end{align}
where $D(t,t')$ is the dissipation kernel.

With these definitions, we can compute the thermodynamic functionals associated with the coherent and squeezed interaction. The stochastic work, Eq.~\ref{eq:work def} performed by the driving force is
\begin{equation}
    W[x] = -2 F_0 |\alpha| e^{-r} \int_0^\tau dt \, e^{-\gamma t} \left[ \gamma \cos(\omega_c t) + \omega_c \sin(\omega_c t) \right] x(t).
\end{equation}
The heat is then derived from the first law of stochastic thermodynamics (Eq.~\ref{eq:heat_def})
\begin{equation}
    Q[x] = \Delta \mathcal{H} - W[x] = \Delta K + \Delta V - \left[ f(\tau)x(\tau) - f(0)x(0) \right] - W[x].
\end{equation}
Here, $\Delta K$ and $\Delta V$ denote the changes in kinetic and potential energy, respectively. These are stochastic quantities due to the randomness inherent in the initial and final phase-space configurations. For instance, even if the particle is initially prepared in a specific quantum state (e.g., a Fock state), the semiclassical description captures the inherent quantum uncertainty through the initial Wigner distribution, which manifests as noise in the classical trajectories. This framework allows us to compute the energy exchange circumventing the complexities of a fully quantum mechanical description.

It is worth noting that for $r=0$ (no squeezing), the non-stationary noise component vanishes ($\sinh(0)=0$), and the system strictly obeys the FDT. However, even in the presence of squeezing ($r \neq 0$), the stationary component of the noise remains proportional to the dissipation kernel. This allows us to identify a `partial' FDT, where $K_0$ represents the stationary kernel complying with the fluctuation-dissipation relation, while $\mathcal{V}$ represents the non-stationary perturbation.

Specifically, we decompose the total noise kernel as $K(t,t') = K_0(t,t') + \mathcal{V}(t,t')$, with
$ K_0(t,t') = \langle \eta_{\text{st}}(t) \eta_{\text{st}}(t') \rangle$, $\mathcal{V}(t,t') = \langle \eta_{\text{n-st}}(t) \eta_{\text{n-st}}(t') \rangle$. The stationary part satisfies a modified FDT with a rescaled effective temperature
\begin{equation}
     K_0(t,t') = \frac{2\cosh(2r)}{\beta} D(t,t').
\end{equation}

Finally, the irreversibility (containing the entropy production due to heat plus the additional contribution from squeezing) is given by
\begin{equation}
  \Delta \mathcal{I} \approx \frac{\beta}{\cosh(2r)} Q[x] - \frac{\beta}{\cosh(2r)} \int_0^\tau dt \int_0^\tau dt' \int_0^\tau du \, F_N(t) \left[ \mathcal{V}(u,t)K_0^{-1}(u,t') \right] \dot{x}(t').
\end{equation}

In the high-finesse cavity limit, where $\omega_c \gg \gamma$, the inverse of the stationary noise kernel can be approximated by
\begin{equation}
    K_0^{-1}(t,t') \approx \frac{\beta \hbar \omega_c}{2 \cosh(2r) F_0^2} \left[\frac{\omega_c^2}{\gamma} \left( \omega_c e^{-\omega_c |t-t'|} - \frac{3}{2} \delta(t-t') \right) - \frac{1}{2\gamma}\delta''(t-t')\right],
\end{equation}
as derived in Appendix~\ref{kernels app}. While the explicit analytical evaluation of the convolution integral involving the non-stationary perturbation $\mathcal{V}$ is possible, the resulting expression is lengthy. For the sake of brevity, we refer the reader to Appendix~\ref{kernels app} for the full mathematical details.

The crucial outcome of this analysis is the modified form of the Second Law, which reads
\begin{equation}
    \langle \Delta s_{\text{sys}} \rangle + \frac{\beta}{\cosh(2r)} \langle Q[x] \rangle \geq \frac{\beta}{\cosh(2r)} \int_0^\tau dt \int_0^\tau dt' \, \langle F_N(t) \dot{x}(t') \rangle \mathcal{G}(t,t').
\end{equation}
Here, $\mathcal{G}(t, t')$ represents the convolution between the non-stationary squeezing kernel and the inverse stationary kernel that is given in the Appendix~\ref{kernels app} .

Two important physical consequences arise from this inequality. First, the prefactor $\frac{\beta}{2\cosh(2r)}$ suggests that the squeezing effectively rescales the bath temperature to a higher effective value, $T_{\text{eff}} \propto T \cosh(2r)$. This implies that as squeezing increases, the relative weight of the standard heat exchange in the entropy balance diminishes. Second, and more crucially, the term on the right-hand side (emerging from the convolution $\mathcal{G}$) can assume negative values. This negative contribution lowers the thermodynamic bound, effectively permitting transient apparent violations of the standard Second Law ($\langle \Delta s_{\text{sys}} \rangle + \beta \langle Q \rangle < 0$).

In our framework, this phenomenon admits a transparent physical interpretation: the non-stationary correlations of the squeezed bath act as a thermodynamic resource, effectively serving as a source of order (since it diminishes the entropy). The consumption of this resource allows for an entropy reduction in the system that would be strictly forbidden in standard thermal environments. Crucially, while such effects are well-established in full quantum thermodynamics, they are typically absent in standard stochastic thermodynamics descriptions. Our formalism bridges this gap, demonstrating its distinct utility by capturing quantum thermodynamic resources within a semiclassical stochastic framework.

\section{Conclusion}

In this work, we have demonstrated that stochastic thermodynamics can be extended to describe classical degrees of freedom coupled to a quantum environment. The resulting Quantum Induced Stochastic Dynamics reveals that the environment induces colored and potentially non-thermal noise, contingent upon the specific quantum state traced out. This formalism unveils a rich variety of dynamical regimes accessible to the classical system, determined by the nature of the quantum environment.

Due to the quantum nature of the environment, we reinterpreted the physics of energy exchange in this hybrid setting. We identified work with the ordered energy transfer that drives the particle's motion, while heat is associated with the disordered energy exchange arising from quantum decoherence and dissipation. This perspective distinguishes our approach from standard stochastic thermodynamics: here, the quantum environment acts as a unified dynamic bath capable of simultaneously performing work and exchanging heat, in contrast to the standard tripartite separation (system, external agent, and passive reservoir) typically found in the stochastic thermodynamics framework.

We further established that when the environment satisfies the Fluctuation-Dissipation Theorem, the path entropy—quantifying irreversibility—is directly linked to the dissipated heat. However, for non-equilibrium quantum states, such as squeezed thermal states, this simple relation breaks down. We derived a modified Second Law that incorporates an additional term accounting for the non-stationary correlations. This term reflects the consumption of quantum resources (squeezing) and allows for apparent negative entropy production rates in the system, consistent with the thermodynamic cost of maintaining the quantum state. Thus, the stochastic thermodynamics of QISD framework connects quantum resources and classical stochastic dynamics.

Finally, the framework presented here enables investigations into more complex systems, particularly where multiplicative noise emerges directly from the interaction, as seen in gravitational wave detection contexts \cite{parikh2021signatures}. Moreover, our results are particularly relevant for levitated optomechanics, where the nanoparticle operates as a classical degree of freedom coupled to a quantum state of light \cite{paraguassu2024quantum}. This configuration offers a ground for engineering specific noise properties from quantum states of light to optimize thermodynamic protocols and heat engines.

\section*{Acknowledges}
We acknowledge Daniel Martinez-Tibaduiza, Pedro Barreto, Rui Aquino, Felipe Sobrero, Bruno Suassuna, Gustavo Forão, Luca Abrahão, Thiago Guerreiro, Welles Morgado for useful discussions. P.V.P acknowledges the Funda\c{c}\~ao de Amparo \`a Pesquisa do Estado do Rio de Janeiro (FAPERJ Process SEI-260003/000174/2024). This work was partially supported by StoneLab.

\bibliography{name}

@article{o2010quantum,
  title={Quantum ground state and single-phonon control of a mechanical resonator},
  author={O’Connell, Aaron D and Hofheinz, Max and Ansmann, Markus and Bialczak, Radoslaw C and Lenander, Mike and Lucero, Erik and Neeley, Matthew and Sank, Daniel and Wang, H and Weides, Martin and others},
  journal={Nature},
  volume={464},
  number={7289},
  pages={697--703},
  year={2010},
  publisher={Nature Publishing Group UK London}
}

@article{tebbenjohanns2021quantum,
  title={Quantum control of a nanoparticle optically levitated in cryogenic free space},
  author={Tebbenjohanns, Felix and Mattana, M Luisa and Rossi, Massimiliano and Frimmer, Martin and Novotny, Lukas},
  journal={Nature},
  volume={595},
  number={7867},
  pages={378--382},
  year={2021},
  publisher={Nature Publishing Group UK London}
}

@article{darabi2023stochastic,
  title={Stochastic energetics of a colloidal particle trapped in a viscoelastic bath},
  author={Darabi, Farshad and Ferrer, Brandon R and Gomez-Solano, Juan Ruben},
  journal={New Journal of Physics},
  volume={25},
  number={10},
  pages={103021},
  year={2023},
  publisher={IOP Publishing}
}

@article{guevara2023brownian,
  title={A Brownian cyclic engine operating in a viscoelastic active suspension},
  author={Guevara-Valadez, Carlos Antonio and Marathe, Rahul and Gomez-Solano, Juan Ruben},
  journal={Physica A: Statistical Mechanics and its Applications},
  volume={609},
  pages={128342},
  year={2023},
  publisher={Elsevier}
}

@misc{camurati2026,
  author        = {Camurati, Antonio and Sobrero, Felipe and Suassuna, Bruno and Paraguass{\'u}, Pedro V.},
  title         = {From Feynman-Vernon to Wiener Stochastic Path Integral},
  year          = {2026},
  eprint        = {2602.00258},
  archivePrefix = {arXiv},
  primaryClass  = {quant-ph},
  url           = {https://arxiv.org/abs/2602.00258}
}

@article{artini2025non,
  title={Non-equilibrium thermodynamics of the quantum Brownian motion: Anomalous non-equilibrium currents arising from complete positivity},
  author={Artini, Simone and Monaco, Gabriele Lo and Imparato, Alberto and Paternostro, Mauro and Donadi, Sandro},
  journal={arXiv preprint arXiv:2507.23322},
  year={2025}
}

@article{kremer2024all,
  title={All-electrical cooling of an optically levitated nanoparticle},
  author={Kremer, Oscar and Califrer, Igor and Tandeitnik, Daniel and von der Weid, Jean Pierre and Tempor{\~a}o, Guilherme and Guerreiro, Thiago},
  journal={Physical Review Applied},
  volume={22},
  number={2},
  pages={024010},
  year={2024},
  publisher={APS}
}

@article{delic2020cooling,
  title={Cooling of a levitated nanoparticle to the motional quantum ground state},
  author={Deli{\'c}, Uro{\v{s}} and Reisenbauer, Manuel and Dare, Kahan and Grass, David and Vuleti{\'c}, Vladan and Kiesel, Nikolai and Aspelmeyer, Markus},
  journal={Science},
  volume={367},
  number={6480},
  pages={892--895},
  year={2020},
  publisher={American Association for the Advancement of Science}
}

@article{piotrowski2023simultaneous,
  title={Simultaneous ground-state cooling of two mechanical modes of a levitated nanoparticle},
  author={Piotrowski, Johannes and Windey, Dominik and Vijayan, Jayadev and Gonzalez-Ballestero, Carlos and de los R{\'\i}os Sommer, Andr{\'e}s and Meyer, Nadine and Quidant, Romain and Romero-Isart, Oriol and Reimann, Ren{\'e} and Novotny, Lukas},
  journal={Nature Physics},
  volume={19},
  number={7},
  pages={1009--1013},
  year={2023},
  publisher={Nature Publishing Group UK London}
}

@article{chan2011laser,
  title={Laser cooling of a nanomechanical oscillator into its quantum ground state},
  author={Chan, Jasper and Alegre, TP Mayer and Safavi-Naeini, Amir H and Hill, Jeff T and Krause, Alex and Gr{\"o}blacher, Simon and Aspelmeyer, Markus and Painter, Oskar},
  journal={Nature},
  volume={478},
  number={7367},
  pages={89--92},
  year={2011},
  publisher={Nature Publishing Group UK London}
}

@article{bild2023schrodinger,
  title={Schr{\"o}dinger cat states of a 16-microgram mechanical oscillator},
  author={Bild, Marius and Fadel, Matteo and Yang, Yu and Von L{\"u}pke, Uwe and Martin, Phillip and Bruno, Alessandro and Chu, Yiwen},
  journal={Science},
  volume={380},
  number={6642},
  pages={274--278},
  year={2023},
  publisher={American Association for the Advancement of Science}
}

@article{ockeloen2018stabilized,
  title={Stabilized entanglement of massive mechanical oscillators},
  author={Ockeloen-Korppi, CF and Damsk{\"a}gg, E and Pirkkalainen, J-M and Asjad, M and Clerk, AA and Massel, F and Woolley, MJ and Sillanp{\"a}{\"a}, MA},
  journal={Nature},
  volume={556},
  number={7702},
  pages={478--482},
  year={2018},
  publisher={Nature Publishing Group UK London}
}

@article{parikh2021signatures,
  title={Signatures of the quantization of gravity at gravitational wave detectors},
  author={Parikh, Maulik and Wilczek, Frank and Zahariade, George},
  journal={Physical Review D},
  volume={104},
  number={4},
  pages={046021},
  year={2021},
  publisher={APS}
}

@article{parikh2025quantum,
  title={Quantum-gravitational noise correlation in nearby detectors},
  author={Parikh, Maulik and Setti, Francesco},
  journal={Physical Review D},
  volume={111},
  number={4},
  pages={046004},
  year={2025},
  publisher={APS}
}

@article{paraguassu2024quantum,
  title={Quantum-induced stochastic optomechanical dynamics},
  author={Paraguass{\'u}, Pedro V and Abrah{\~a}o, Luca and Guerreiro, Thiago},
  journal={Journal of the Optical Society of America B},
  volume={41},
  number={12},
  pages={2798--2810},
  year={2024},
  publisher={Optica Publishing Group}
}

@article{paraguassu2025apparent,
  title={Apparent violations of the second law in the quantum-classical dynamics of interacting levitated nanoparticles},
  author={Paraguass{\'u}, Pedro V and Guerreiro, Thiago},
  journal={Physical Review E},
  volume={112},
  number={2},
  pages={024134},
  year={2025},
  publisher={APS}
}

@article{sobrero2025response,
  title={Response of a classical mesoscopic oscillator to a two-level quantum system},
  author={Sobrero, Felipe and Abrah{\~a}o, Luca and Guerreiro, Thiago and Paraguass{\'u}, Pedro V},
  journal={arXiv preprint arXiv:2509.04216},
  year={2025}
}

@article{seifert2005entropy,
  title={Entropy production along a stochastic trajectory and an integral fluctuation theorem},
  author={Seifert, Udo},
  journal={Physical review letters},
  volume={95},
  number={4},
  pages={040602},
  year={2005},
  publisher={APS}
}

@article{chernyak2006path,
  title={Path-integral analysis of fluctuation theorems for general Langevin processes},
  author={Chernyak, Vladimir Y and Chertkov, Michael and Jarzynski, Christopher},
  journal={Journal of Statistical Mechanics: Theory and Experiment},
  volume={2006},
  number={08},
  pages={P08001},
  year={2006},
  publisher={IOP Publishing}
}

@article{melo2024brownian,
  title={Brownian Fluctuations of a non-confining potential},
  author={Melo, Pedro B and Paraguass{\'u}, Pedro V and Nascimento, Eduardo S and Morgado, Welles AM},
  journal={Physica A: Statistical Mechanics and its Applications},
  volume={650},
  pages={129996},
  year={2024},
  publisher={Elsevier}
}

@article{paraguassu2024brownian,
  title={Brownian fluctuations of kinetic energy under Lorentz force},
  author={Paraguass{\'u}, Pedro V},
  journal={Journal of Statistical Mechanics: Theory and Experiment},
  volume={2024},
  number={1},
  pages={013202},
  year={2024},
  publisher={IOP Publishing}
}

@book{wio2013path,
  title={Path integrals for stochastic processes: An introduction},
  author={Wio, Horacio S},
  year={2013},
  publisher={World Scientific}
}

@book{weiss2012quantum,
  title={Quantum dissipative systems},
  author={Weiss, Ulrich},
  year={2012},
  publisher={World Scientific}
}

@article{feynman2000theory,
  title={The theory of a general quantum system interacting with a linear dissipative system},
  author={Feynman, Richard Phillips and Vernon Jr, Frank L},
  journal={Annals of physics},
  volume={281},
  number={1-2},
  pages={547--607},
  year={2000},
  publisher={Elsevier}
}

@book{kamenev2023field,
  title={Field theory of non-equilibrium systems},
  author={Kamenev, Alex},
  year={2023},
  publisher={Cambridge University Press}
}

@article{caldeira1983path,
  title={Path integral approach to quantum Brownian motion},
  author={Caldeira, Amir O and Leggett, Anthony J},
  journal={Physica A: Statistical mechanics and its Applications},
  volume={121},
  number={3},
  pages={587--616},
  year={1983},
  publisher={Elsevier}
}

@article{calzetta1994noise,
  title={Noise and fluctuations in semiclassical gravity},
  author={Calzetta, Esteban and Hu, BL},
  journal={Physical Review D},
  volume={49},
  number={12},
  pages={6636},
  year={1994},
  publisher={APS}
}

@article{seifert2012stochastic,
  title={Stochastic thermodynamics, fluctuation theorems and molecular machines},
  author={Seifert, Udo},
  journal={Reports on progress in physics},
  volume={75},
  number={12},
  pages={126001},
  year={2012},
  publisher={IOP Publishing}
}

@article{jarzynski2011equalities,
  author = {Jarzynski, Christopher},
  title = {Equalities and Inequalities: Irreversibility and the Second Law of Thermodynamics at the Nanoscale},
  journal = {Annual Review of Condensed Matter Physics},
  volume = {2},
  number = {1},
  pages = {329-351},
  year = {2011},
  doi = {10.1146/annurev-conmatphys-062910-140506},
  URL = {https://doi.org/10.1146/annurev-conmatphys-062910-140506}
}

@article{jarzynski1997nonequilibrium,
  title={Nonequilibrium equality for free energy differences},
  author={Jarzynski, Christopher},
  journal={Physical Review Letters},
  volume={78},
  number={14},
  pages={2690},
  year={1997},
  publisher={APS}
}

@article{sekimoto1998langevin,
  title={Langevin equation and thermodynamics},
  author={Sekimoto, Ken},
  journal={Progress of Theoretical Physics Supplement},
  volume={130},
  pages={17--27},
  year={1998},
  publisher={Oxford Academic}
}

@article{collin2005verification,
  title={Verification of the Crooks fluctuation theorem and recovery of RNA folding free energies},
  author={Collin, Delphine and Ritort, Felix and Jarzynski, Christopher and Smith, Steven B and Tinoco Jr, Ignacio and Bustamante, Carlos},
  journal={Nature},
  volume={437},
  number={7056},
  pages={231--234},
  year={2005},
  publisher={Nature Publishing Group UK London}
}

@article{alemany2012experimental,
  title={Experimental free-energy measurements of kinetic molecular states using fluctuation theorems},
  author={Alemany, Anna and Mossa, Alessandro and Junier, Ivan and Ritort, Felix},
  journal={Nature Physics},
  volume={8},
  number={9},
  pages={688--694},
  year={2012},
  publisher={Nature Publishing Group UK London}
}

@article{martinez2016brownian,
  title={Brownian carnot engine},
  author={Mart{\'\i}nez, Ignacio A and Rold{\'a}n, {\'E}dgar and Dinis, Luis and Petrov, Dmitri and Parrondo, Juan MR and Rica, Ra{\'u}l A},
  journal={Nature physics},
  volume={12},
  number={1},
  pages={67--70},
  year={2016},
  publisher={Nature Publishing Group UK London}
}

@article{holubec2021fluctuations,
  title={Fluctuations in heat engines},
  author={Holubec, Viktor and Ryabov, Artem},
  journal={Journal of Physics A: Mathematical and Theoretical},
  volume={55},
  number={1},
  pages={013001},
  year={2021},
  publisher={IOP Publishing}
}

@article{li2024realization,
  title={Realization of an all-optical underdamped stochastic Stirling engine},
  author={Li, Chuang and Zhu, Shaochong and He, Peitong and Wang, Yingying and Zheng, Yi and Zhang, Kexin and Gao, Xiaowen and Dong, Ying and Hu, Huizhu},
  journal={Physical Review A},
  volume={109},
  number={2},
  pages={L021502},
  year={2024},
  publisher={APS}
}

@article{forao2025statistics,
  title={Statistics of power and efficiency for collisional Brownian engines},
  author={For{\~a}o, Gustavo AL and Filho, Fernando S and Paraguass{\'u}, Pedro V},
  journal={Physical Review E},
  volume={112},
  number={2},
  pages={024110},
  year={2025},
  publisher={APS}
}

@article{manosas2007force,
  title={Force unfolding kinetics of RNA using optical tweezers. II. Modeling experiments},
  author={Manosas, Maria and Wen, J-D and Li, Pan TX and Smith, Steven B and Bustamante, Carlos and Tinoco, Ignacio and Ritort, Felix},
  journal={Biophysical journal},
  volume={92},
  number={9},
  pages={3010--3021},
  year={2007},
  publisher={Elsevier}
}

@article{tao2024general,
  title={General relativistic stochastic thermodynamics},
  author={Tao, Wang and Cai, Yifan and Cui, Long and Liu, Zhao},
  journal={SciPost Physics Core},
  volume={7},
  number={4},
  pages={082},
  year={2024}
}

@article{cai2025fluctuation,
  title={Fluctuation theorems in general relativistic stochastic thermodynamics},
  author={Cai, Yifan and Wang, Tao and Zhao, Liu},
  journal={Physical Review E},
  volume={111},
  number={2},
  pages={024102},
  year={2025},
  publisher={APS}
}

@article{cai2023relativistic,
  title={Relativistic stochastic mechanics I: Langevin equation from observer’s perspective},
  author={Cai, Yifan and Wang, Tao and Zhao, Liu},
  journal={Journal of Statistical Physics},
  volume={190},
  number={12},
  pages={193},
  year={2023},
  publisher={Springer}
}

@article{cai2023relativistic2,
  title={Relativistic stochastic mechanics II: Reduced Fokker-Planck equation in curved spacetime},
  author={Cai, Yifan and Wang, Tao and Zhao, Liu},
  journal={arXiv preprint arXiv:2307.07805},
  year={2023}
}

@article{pal2020stochastic,
  title={Stochastic thermodynamics of relativistic Brownian motion},
  author={Pal, PS and Deffner, Sebastian},
  journal={New Journal of Physics},
  volume={22},
  number={7},
  pages={073054},
  year={2020},
  publisher={IOP Publishing}
}

@article{paraguassu2021heat,
  title={Heat distribution of relativistic Brownian motion},
  author={Paraguass{\'u}, Pedro V and Morgado, Welles AM},
  journal={The European Physical Journal B},
  volume={94},
  number={10},
  pages={197},
  year={2021},
  publisher={Springer}
}

@article{koide2011thermodynamic,
  title={Thermodynamic laws and equipartition theorem in relativistic Brownian motion},
  author={Koide, T and Kodama, T},
  journal={Physical Review E—Statistical, Nonlinear, and Soft Matter Physics},
  volume={83},
  number={6},
  pages={061111},
  year={2011},
  publisher={APS}
}

@article{pei2025promoting,
  title={Promoting fluctuation theorems into covariant forms},
  author={Pei, Ji-Hui and Chen, Jin-Fu and Quan, HT},
  journal={Physical Review Letters},
  volume={134},
  number={23},
  pages={237102},
  year={2025},
  publisher={APS}
}

@article{wolpert2024stochastic_perspective,
  author = {Wolpert, David H. and Korbel, Jan and Lynn, Christopher W. and Tasnim, Farita and Grochow, Joshua A. and Kardes, G{\"u}lce and Aimone, James B. and Balasubramanian, Vijay and De Giuli, Eric and Doty, David and Freitas, Nahuel and Marsili, Matteo and Ouldridge, Thomas E. and Richa, Andr{\'e}a W. and Riechers, Paul and Rold{\'a}n, {\'E}dgar and Rubenstein, Brenda and Toroczkai, Zoltan and Paradiso, Joseph},
  title = {Is stochastic thermodynamics the key to understanding the energy costs of computation?},
  journal = {Proceedings of the National Academy of Sciences},
  volume = {121},
  number = {45},
  pages = {e2321112121},
  year = {2024},
  doi = {10.1073/pnas.2321112121},
  URL = {https://www.pnas.org/doi/10.1073/pnas.2321112121}
}

@article{wolpert2019stochastic,
  author={Wolpert, David H.},
  title={The stochastic thermodynamics of computation},
  journal={Journal of Physics A: Mathematical and Theoretical},
  volume={52},
  number={19},
  pages={193001},
  year={2019},
  month={apr},
  publisher={IOP Publishing},
  doi={10.1088/1751-8121/ab0850},
  url={https://dx.doi.org/10.1088/1751-8121/ab0850}
}

@article{helms2025stochastic,
  title = {Stochastic thermodynamic bounds on logical circuit operation},
  author = {Helms, Phillip and Chen, Songela W. and Limmer, David T.},
  journal = {Physical Review E},
  volume = {111},
  issue = {3},
  pages = {034110},
  numpages = {15},
  year = {2025},
  month = {Mar},
  publisher = {American Physical Society},
  doi = {10.1103/PhysRevE.111.034110},
  url = {https://link.aps.org/doi/10.1103/PhysRevE.111.034110}
}

@article{tome2023stochastic,
  title={Stochastic thermodynamics of opinion dynamics models},
  author={Tom{\'e}, T{\^a}nia and Fiore, Carlos E and de Oliveira, M{\'a}rio J},
  journal={Physical Review E},
  volume={107},
  number={6},
  pages={064135},
  year={2023},
  publisher={APS}
}

@article{hawthorne2023nonequilibrium,
  title={Nonequilibrium thermodynamics of the majority vote model},
  author={Hawthorne, Felipe and Harunari, Pedro E and de Oliveira, M{\'a}rio J and Fiore, Carlos E},
  journal={Entropy},
  volume={25},
  number={8},
  pages={1230},
  year={2023},
  publisher={MDPI}
}

@article{koide2025unification,
  title={Unification of stochastic and quantum thermodynamics in scalar field theory via a model with a Brownian thermostat},
  author={Koide, T and Nicacio, F},
  journal={Physical Review E},
  volume={112},
  number={2},
  pages={024127},
  year={2025},
  publisher={APS}
}

@article{binder2015quantum,
  title={Quantum thermodynamics of general quantum processes},
  author={Binder, Felix and Vinjanampathy, Sai and Modi, Kavan and Goold, John},
  journal={Physical Review E},
  volume={91},
  number={3},
  pages={032119},
  year={2015},
  publisher={APS}
}

@article{strasberg2019operational,
  title={Operational approach to quantum stochastic thermodynamics},
  author={Strasberg, Philipp},
  journal={Physical Review E},
  volume={100},
  number={2},
  pages={022127},
  year={2019},
  publisher={APS}
}

@article{elouard2017role,
  title={The role of quantum measurement in stochastic thermodynamics},
  author={Elouard, Cyril and Herrera-Mart{\'\i}, David A and Clusel, Maxime and Auff{\`e}ves, Alexia},
  journal={npj Quantum Information},
  volume={3},
  number={1},
  pages={9},
  year={2017},
  publisher={Nature Publishing Group UK London}
}

@article{dabelow2019irreversibility,
  title={Irreversibility in active matter systems: Fluctuation theorem and mutual information},
  author={Dabelow, Lennart and Bo, Stefano and Eichhorn, Ralf},
  journal={Physical Review X},
  volume={9},
  number={2},
  pages={021009},
  year={2019},
  publisher={APS}
}

@article{bebon2025thermodynamics,
  title={Thermodynamics of active matter: Tracking dissipation across scales},
  author={Bebon, Robin and Robinson, Joshua F and Speck, Thomas},
  journal={Physical Review X},
  volume={15},
  number={2},
  pages={021050},
  year={2025},
  publisher={APS}
}

@article{caprini2019entropy,
  title={The entropy production of Ornstein--Uhlenbeck active particles: a path integral method for correlations},
  author={Caprini, Lorenzo and Marconi, Umberto Marini Bettolo and Puglisi, Andrea and Vulpiani, Angelo},
  journal={Journal of Statistical Mechanics: Theory and Experiment},
  volume={2019},
  number={5},
  pages={053203},
  year={2019},
  publisher={IOP Publishing}
}

@article{paraguassu2025effects,
  title={Effects of kinetic energy on heat fluctuations of passive and active overdamped driven particles},
  author={Paraguass{\'u}, Pedro V and Aquino, Rui and de Castro, Pablo},
  journal={Physical Review E},
  volume={111},
  number={3},
  pages={034111},
  year={2025},
  publisher={APS}
}

@article{funo2018path,
  title={Path integral approach to quantum thermodynamics},
  author={Funo, Ken and Quan, HT},
  journal={Physical review letters},
  volume={121},
  number={4},
  pages={040602},
  year={2018},
  publisher={APS}
}

@book{zwanzig2001nonequilibrium,
  title={Nonequilibrium statistical mechanics},
  author={Zwanzig, Robert},
  year={2001},
  publisher={Oxford university press}
}

@book{peliti2021stochastic,
  title={Stochastic thermodynamics: an introduction},
  author={Peliti, Luca and Pigolotti, Simone},
  year={2021},
  publisher={Princeton University Press}
}

@book{sekimoto2010stochastic,
  title={Stochastic Energetics},
  author={Sekimoto, Ken},
  volume={799},
  year={2010},
  publisher={Springer},
  address={Berlin, Heidelberg},
  doi={10.1007/978-3-642-05411-2},
  isbn={978-3-642-05410-5}
}

@book{seifert2025stochastic,
  title={Stochastic thermodynamics},
  author={Seifert, Udo},
  year={2025},
  publisher={CAMBRIDGE University Press}
}

@article{pires2023optimal,
  title={Optimal time-entropy bounds and speed limits for Brownian thermal shortcuts},
  author={Pires, Lu{\'\i}s Barbosa and Goerlich, R{\'e}mi and da Fonseca, Arthur Luna and Debiossac, Maxime and Hervieux, Paul-Antoine and Manfredi, Giovanni and Genet, Cyriaque},
  journal={Physical Review Letters},
  volume={131},
  number={9},
  pages={097101},
  year={2023},
  publisher={APS}
}

@article{jun2014high,
  title={High-precision test of Landauer’s principle in a feedback trap},
  author={Jun, Yonggun and Gavrilov, Mom{\v{c}}ilo and Bechhoefer, John},
  journal={Physical review letters},
  volume={113},
  number={19},
  pages={190601},
  year={2014},
  publisher={APS}
}

@article{sagawa2012nonequilibrium,
  title={Nonequilibrium thermodynamics of feedback control},
  author={Sagawa, Takahiro and Ueda, Masahito},
  journal={Physical Review E—Statistical, Nonlinear, and Soft Matter Physics},
  volume={85},
  number={2},
  pages={021104},
  year={2012},
  publisher={APS}
}

@article{searles1999fluctuation,
  title={Fluctuation theorem for stochastic systems},
  author={Searles, Debra J and Evans, Denis J},
  journal={Physical Review E},
  volume={60},
  number={1},
  pages={159},
  year={1999},
  publisher={APS}
}

@article{evans2002fluctuation,
  title={The fluctuation theorem},
  author={Evans, Denis J and Searles, Debra J},
  journal={Advances in Physics},
  volume={51},
  number={7},
  pages={1529--1585},
  year={2002},
  publisher={Taylor \& Francis}
}

@book{itzykson2006quantum,
  title={Quantum Field Theory},
  author={Itzykson, Claude and Zuber, Jean-Bernard},
  year={2006},
  publisher={Dover Publications},
  address={Mineola, New York},
  note={Chapter 1 and Appendix on Integral Equations}
}

@article{cohen2008properties,
  title={Properties of nonequilibrium steady states: a path integral approach},
  author={Cohen, EGD},
  journal={Journal of Statistical Mechanics: Theory and Experiment},
  volume={2008},
  number={07},
  pages={P07014},
  year={2008},
  publisher={IOP Publishing}
}

@book{fox2006quantum,
  title={Quantum Optics: An Introduction},
  author={Fox, Mark},
  volume={15},
  year={2006},
  publisher={Oxford University Press},
  address={New York, USA},
  series={Oxford Master Series in Physics},
  isbn={978-0198566731}
}

@article{gonzalez2023suppressing,
  title={Suppressing recoil heating in levitated optomechanics using squeezed light},
  author={Gonzalez-Ballestero, Carlos and Zieli{\'n}ska, Joanna A and Rossi, Massimiliano and Militaru, Andrei and Frimmer, Martin and Novotny, Lukas and Maurer, Patrick and Romero-Isart, Oriol},
  journal={PRX Quantum},
  volume={4},
  number={3},
  pages={030331},
  year={2023},
  publisher={APS}
}

@article{delic2019cavity,
  title={Cavity cooling of a levitated nanosphere by coherent scattering},
  author={Deli{\'c}, Uro{\v{s}} and Reisenbauer, Manuel and Grass, David and Kiesel, Nikolai and Vuleti{\'c}, Vladan and Aspelmeyer, Markus},
  journal={Physical review letters},
  volume={122},
  number={12},
  pages={123602},
  year={2019},
  publisher={APS}
}

\appendix

\section{Time Reversal Analysis}
\label{app:time_reversal}

In this appendix, we analyze the operation of time reversal applied to the equation of motion given by
\begin{equation}
    \tilde{\mathcal{E}}[\tilde{x}] = m \ddot{\tilde{x}} + V'(\tilde{x}) - \tilde{F}_{\text{diss}}[\tilde{x}] - \tilde{F}_{\text{det}}(t).
\end{equation}
Under time reversal, the coordinates transform as $\tilde{x}(t)= x(\tau-t)$ and the deterministic force as $\tilde{F}_{\text{det}}(t)= F_{\rm det}(\tau-t)$. The second derivative is invariant under this transformation. Thus, we focus on the dissipation term.

We consider a time $\bar{t}$ in the reversed frame (which corresponds to the forward time $t$) and evaluate the memory integral from this instant up to the final time $\tau$. The calculation proceeds as follows:
\begin{align}
    \tilde{F}_{\text{diss}}(\bar{t}) &= -\int_{\bar{t}}^{\tau} D(s'-\bar{t}) \dot{\tilde{x}}(s') ds' \nonumber \\
    &= \int_{\bar{t}}^{\tau} D(s'-\bar{t}) \dot{x}(\tau-s') ds' \nonumber \\
    &= -\int_{t}^{0} D(\tau-u-\bar{t}) \dot{x}(u) du \nonumber \\
    &= \int_0^t D(t-u) \dot{x}(u) du.
    \label{eq:diss_calc}
\end{align}
In the derivation above, we first employed the time-reversal relation $\dot{\tilde{x}}(s') = -\dot{x}(\tau-s')$ in the second line, which introduces a sign change. Subsequently, in the third line, we performed the change of variables $u = \tau - s'$, implying $ds' = -du$. Under this transformation, the integration limits map as follows: the lower limit $s'=\bar{t}$ corresponds to $u=\tau-\bar{t} \equiv t$, and the upper limit $s'=\tau$ corresponds to $u=0$. Finally, in the fourth line, we reversed the integration limits ($\int_t^0 \to -\int_0^t$) to absorb the remaining negative sign from the differential, thereby recovering the standard form of the dissipation term.

\section{Path entropy calculation}
\label{path entropy app}

In this appendix, we detail the derivation of the path entropy production, specifically focusing on the manipulation of the double time integrals. We start by expanding the expression for the irreversibility $\Delta \mathcal{I}$, which involves the noise kernel $K^{-1}(t,t')$ and the difference between the reversed and forward dissipation forces.

First, we recall from Appendix \ref{app:time_reversal} that the time-reversed dissipation force transforms in such a way that its structure mirrors the forward case but with inverted limits. Consequently, the difference term becomes additive
\begin{equation}
    \tilde{F}_{\text{diss}}[\tilde{x}] - F_{\text{diss}}[x] = 2 \int_0^t ds D(t,s) \dot{x}(s).
\end{equation}
Substituting this result into the expression for $\Delta \mathcal{I}$, we obtain a triple integral structure
\begin{align}
    \Delta \mathcal{I} &= \int_0^\tau dt \int_0^\tau dt' F_N(t') K^{-1}(t,t')\left[\tilde{F}_{\text{diss}} - F_{\text{diss}}\right] \nonumber \\
    &= 2 \int_0^\tau dt \int_0^\tau dt' F_N(t') K^{-1}(t,t') \int_0^t ds D(t,s) \dot{x}(s).
\end{align}
To identify the effective kernel acting on the velocity $\dot{x}(t)$, we must rearrange the order of integration between $t$ and $s$. Note that the current integration domain over $t$ and $s$ defines a triangular region $0 \le s \le t \le \tau$. To identify the effective kernel acting on the velocity $\dot{x}(t)$, we must rearrange the double integral. We perform this manipulation by first exchanging the order of integration between $t$ and $s$. Since the original integration domain is defined by the triangular region $0 \le s \le t \le \tau$, switching the order results in the outer integral running over $s$ from $0$ to $\tau$, while the inner integral runs over $t$ from $s$ to $\tau$. Following this, we relabel the dummy variables ($s \to t$ and $t \to s$) to explicitly express the velocity term as $\dot{x}(t)$ in the final equation. This yields
\begin{equation}
    \Delta \mathcal{I} = 2 \int_0^\tau dt \int_0^\tau dt' F_N(t') \dot{x}(t) \underbrace{ \int_t^\tau ds D(s,t) K^{-1}(t',s) }_{\mathcal{K}(t,t')},
\end{equation}
Here, we have isolated the velocity $\dot{x}(t)$ and identified the effective kernel $\mathcal{K}(t,t')$, which convolves the memory kernel $D(s,t)$ with the noise correlation inverse $K^{-1}(t',s)$ over the future interval.

\section{Kernels Calculations}
\label{kernels app}

\subsection{Inverse of the Stationary Kernel $K_0(t,t')$}

To compute the inverse of the noise kernel, we exploit the time-translational invariance of the system. Since the kernel depends only on the time difference $\tau = t - t'$, we can operate in the Fourier domain. The Fourier transform of the kernel is given by
\begin{align}
    \tilde{K}_0(\omega) &= \int_{-\infty}^{\infty} d\tau K_0(\tau) e^{i\omega \tau} \nonumber \\
    &= \frac{2 \cosh(2r) F_0^2}{\beta \hbar \omega_c} \left[\frac{\gamma}{\gamma^2 + (\omega - \omega_0)^2} + \frac{\gamma}{\gamma^2 + (\omega + \omega_0)^2}\right].
\end{align}
The inverse in the Fourier space is simply the reciprocal, $1/\tilde{K}_0(\omega)$. Algebraic manipulation yields
\begin{equation}
    \tilde{K}_0^{-1}(\omega) = \frac{\beta \hbar \omega_c}{2 \cosh(2r) F_0^2} \frac{\left[\gamma^2 + (\omega - \omega_0)^2\right] \left[\gamma^2 + (\omega + \omega_0)^2\right]}{2 \gamma (\gamma^2 + \omega^2 + \omega_0^2)}.
\end{equation}
By performing the inverse Fourier transform back to the time domain, we obtain the exact form of the inverse kernel
\begin{align}
    K_0^{-1}(t,s) = \frac{\beta \hbar \omega_c}{2 \cosh(2r) F_0^2} \Bigg[ \frac{\omega_0^2 \sqrt{\gamma^2 + \omega_0^2}}{\gamma} e^{-\sqrt{\gamma^2 + \omega_0^2} |t-s|} 
     + \left( \frac{\gamma}{2} - \frac{3 \omega_0^2}{2 \gamma} \right) \delta(t-s) - \frac{1}{2 \gamma}\frac{d^2}{dt^2}\delta(t-s) \Bigg].
\end{align}
In the weak coupling regime, where $\omega_0 \gg \gamma$, this expression simplifies significantly. Approximating terms of order $\gamma/\omega_0$, we arrive at
\begin{equation}
    K_0^{-1}(t,s) \approx \frac{\beta \hbar \omega_c}{2 \cosh(2r) F_0^2} \left[\frac{\omega_0^2}{\gamma} \left( \omega_0 e^{-\omega_0 |t-s|} - \frac{3}{2} \delta(t-s) \right) - \frac{1}{2\gamma}\delta''(t-s)\right].
    \label{eq:approx_K}
\end{equation}

\subsection{Calculation of the Effective Kernel $\mathcal{G}(t,t')$}

With the inverse kernel derived, we proceed to compute the combined kernel $\mathcal{G}(t,t')$, which couples the non-stationary potential interaction with the inverse noise correlations. It is defined by the integral
\begin{equation}
    \mathcal{G}(t,t') = \int_0^\tau du \, \mathcal{V}(u,t) K_0^{-1}(u,t').
\end{equation}
Performing the integration directly through symbolic computation and collecting the terms, we express the result as
\begin{widetext}
\begin{equation}
    \mathcal{G}(t,t') = -\frac{\tanh(2r)}{2\nu\omega_0(\gamma^4 + 4\omega_0^4)} 
    \exp\Big[{-\gamma t' - \tau(\gamma + \omega_0) - t(2\gamma + \omega_0)}\Big] 
    \Big[ \mathcal{C}(t,t') + \mathcal{S}(t,t') \Big],
\end{equation}
where the cosine contribution $\mathcal{C}$ and the sine contribution $\mathcal{S}$ are given by
\begin{subequations}
\begin{align}
    \mathcal{C}(t,t') &= -2\omega_0^3(\gamma^3 + \gamma^2\omega_0 - 2\omega_0^3) e^{2t\gamma + \tau(\gamma + \omega_0)} \cos(t'\omega_0) \nonumber \\
    &\quad + \gamma^2(\gamma^4 + 2\gamma^2\omega_0^2 + 8\omega_0^4) e^{(t+\tau)(\gamma + \omega_0)} \cos[(t+t')\omega_0] \nonumber \\
    &\quad + 2\omega_0^3(\gamma^3 - \gamma^2\omega_0 + 2\omega_0^3) e^{2t(\gamma + \omega_0)} \cos[(t'+\tau)\omega_0], \\
    \mathcal{S}(t,t') &= 2\omega_0 \Bigg\{ \omega_0^3(\gamma^2 + 2\gamma\omega_0 + 2\omega_0^2) e^{2t\gamma + \tau(\gamma + \omega_0)} \sin(t'\omega_0) \nonumber \\
    &\quad + \gamma^5 e^{(t+\tau)(\gamma + \omega_0)} \sin[(t+t')\omega_0] \nonumber \\
    &\quad + \omega_0^3(2\gamma\omega_0 - \gamma^2 - 2\omega_0^2) e^{2t(\gamma + \omega_0)} \sin[(t'+\tau)\omega_0] \Bigg\}.
\end{align}
\end{subequations}
\end{widetext}
Depending on the specific time intervals and relative phases, the sign of these contributions may oscillate, reflecting the intricate exchange of energy mediated by the squeezed reservoir.

\end{document}